\documentclass[
 reprint,
 superscriptaddress,
 amsmath,amssymb,
 aps,
 pre,
]{revtex4-2}
\usepackage{graphicx}
\usepackage{dcolumn}
\usepackage{bm}
\usepackage[skip=0.6\baselineskip]{caption}
\usepackage{subcaption}
\usepackage{chngcntr}
 \usepackage{relsize}
\usepackage[colorlinks=true,allcolors=blue]{hyperref}
 
\begin{document}

\title{Stress relaxation in fiber networks via force-dependent stochastic severing}%
\author{Prathamesh Kulkarni}
\affiliation{Department of Chemical \& Biomolecular Engineering, Rice University, Houston, Texas 77005, USA}
\affiliation{Center for Theoretical Biological Physics, Rice University, Houston, Texas 77005, USA}
\author{Anatoly B. Kolomeisky}
\affiliation{Department of Chemistry, Rice University, Houston, Texas 77005, USA}
\affiliation{Center for Theoretical Biological Physics, Rice University, Houston, Texas 77005, USA}
\affiliation{Department of Chemical \& Biomolecular Engineering, Rice University, Houston, Texas 77005, USA}
\affiliation{Department of Physics \& Astronomy, Rice University, Houston, Texas 77005, USA}

\author{Fred C. MacKintosh}
\affiliation{Department of Chemical \& Biomolecular Engineering, Rice University, Houston, Texas 77005, USA}
\affiliation{Center for Theoretical Biological Physics, Rice University, Houston, Texas 77005, USA}
\affiliation{Department of Chemistry, Rice University, Houston, Texas 77005, USA}
\affiliation{Department of Physics \& Astronomy, Rice University, Houston, Texas 77005, USA}

\begin{abstract}
\noindent Fiber networks contribute to the mechanical stability of various biological systems, from cells to tissues. Such systems have been modeled by networks of springs or fibers that exhibit rigidity transitions as a function of either connectivity or applied strain. For a fiber network under constant applied strain, severing can reduce the connectivity and destabilize an initially rigid structure. Here, we investigate stress relaxation in spring and fiber networks in the presence of stochastic, force-dependent severing. A computational model to predict stress relaxation with mechanochemical feedback of stress on severing is developed. We also examine the effects of severing on the network topology and onset of rigidity transition. Using 2D triangular lattice-based computer simulations, we explore different limits of the feedback and demonstrate the shift in the onset of rigidity depending on the limit. The limit of tension-suppressed severing delays stress relaxation and shifts the transition into the bending-dominated regime to lower-than-expected connectivity. In contrast, tension-enhanced severing accelerates relaxation and shifts the transition to higher-than-expected connectivity. It is also found that the magnitude of this shift depends on the applied shear strain and the strength of the feedback. Our theoretical approach clarifies some microscopic aspects of these phenomena. Understanding the impact of such feedback mechanisms can provide valuable insights into designing systems by tuning the feedback to the desired response.
\end{abstract}
\maketitle
\section{INTRODUCTION}
Crosslinked biopolymer networks contribute to the mechanical stability of cells and tissues. These biological systems frequently exhibit a spectrum of phenomena, such as strain-stiffening and rigidity transitions, that are crucial for their functioning.\cite{head2003deformation,WilhelmPRL2003,HeadPRE2003,gardel2004elastic,storm2005nonlinear,onck2005alternative,das2007effective,wyart2008elasticity,licup2015stress,feng2016nonlinear,burla2019mechanical,song2025strain,picu2011mechanics,parvez2023stiffening} 
The mechanical behavior of such networks is strongly influenced by the properties of individual fibers and the network architecture, which is generally characterized by the average coordination number or connectivity ($z$) of the network. Maxwell established a criterion for the marginal stability of mechanical structures, showing that a simple spring network is stable only when the average connectivity exceeds the \textit{isostatic} threshold $z_{c} = 2d$ in the thermodynamic limit, where $d$ represents dimensions \cite{maxwell1864calculation}. 
Physiological biopolymer networks, including extracellular networks such as collagen, have been shown to exhibit typical connectivities of $z\simeq 3.5$ \cite{lindstrom2013finite,stein2011micromechanics,jansen2018role}, placing them well below the isostatic threshold in dimensions 2 and 3. Thus, in the presence of only central force interactions, such networks would be mechanically floppy or unstable at small strains. A stable linear response can be achieved in sub-isostatic networks with non-central-force interactions such as fiber bending \cite{satcher1996theoretical,head2003deformation,WilhelmPRL2003,onck2005alternative,das2007effective,HeadPRE2003}, thermal fluctuations \cite{barsky1996elastic,plischke1998entropic,dennison2013fluctuation,mao2015mechanical,maozhang2016finite,arzash2023finitetemp}, or active stresses \cite{broedersz2011molecular,sheinman2012actively,koenderink2009active,jansen2013cells,chen2020motor,chen2023motor,mizuno2007nonequilibrium}.
 Recent theory and experiments on fiber network behavior have shown a strain-controlled rigidity transition. When subjected to finite incremental shear deformation, the networks transition from floppy to rigid states at a critical strain threshold depending on the network connectivity and geometry. This phase transition exhibits rich critical phenomena and is second-order in nature \cite{sharma2016strain,jansen2018role,arzash2020finite,shivers2019normal,shivers2019scaling,feng2016nonlinear,chen2024field,lerner2023scaling}. 

Network stability can also be affected by fracture when networks such as collagen are subjected to large deformations during injuries or pathological conditions such as aneurysms, where connectivity again plays a key role \cite{burla2020connectivity}. 
Biochemical severing of filaments, such as of F-actin by ADF/cofilin can also affect stability. 
ADF/cofilin cooperatively binds actin filaments to promote depolymerization and severing \cite{enrique2015actin,de2009cofilin,enrique2015mechanical,blanchoin2014actin,pollard2003cellular,mcgough1997cofilin,sun2024cofilin}. The severing rate has been proposed to be influenced by mechanical factors such as tension, torsion, and buckling \cite{pollard1986rate,bugyi2010control,bibeau2023twist,wioland2019torsional}. Filaments in buckled states have been proposed to be easier to sever, while cofilin binding has been observed to be influenced under tension \cite{schramm2017actin,hayakawa2011actin,mccullough2011cofilin}. Filaments found in crosslinked networks are under the exposure of various mechanical stresses, and such fragmentation mechanisms can lead to loss of rigidity. Recent studies have also demonstrated how presence of severing proteins leads to novel stress relaxation behavior that is length-dependent and independent \cite{arzash2019stressrelaxation,mccall2019cofilin}. 

To understand how fragmentation events affect the mechanical response of biopolymer networks, we consider a system of spring networks with force-dependent severing. Fragmentation of filaments in crosslinked networks directly affects the average connectivity and network topology as the system relaxes. Hence, investigating the implications of force-dependent severing on the onset of rigidity in crosslinked fiber networks is important. Recent studies on bond removal, selective pruning, and network design have shown that local changes in network connectivity can strongly influence the global mechanical response \cite{goodrich2015principle,rocks2017designing,hexner2018role,galvani2024building}.

In this work, we focus on understanding the stress relaxation in athermal fiber networks via force-dependent severing. We also investigate the role of network structural properties on the rigidity transition in the presence and absence of mechanochemical feedback of stress on severing. 

We develop a computational model to describe the mechanical behavior of fiber networks in the presence of severing reactions when held under an applied shear strain. Using triangular lattice-based networks, we demonstrate that the mechanical response is in agreement with the rigidity transition phase diagram in the absence of feedback \cite{shivers2019scaling,arzash2020finite}. We introduce force-dependent stochastic severing events and identify different limits of mechanochemical feedback where severing is suppressed, enhanced, and unaffected by tension. We find that stress relaxation is either delayed or occurs earlier depending on the limit. 
We draw analogies for tension-suppressed limits with catch bonds as the fragmentation rate (or ``bond rupture" rate) is reduced by tension. Upon further investigation of stress variation with network connectivity, we find that the onset of rigidity transition shifts depending on the mechanochemical feedback limit. 
We find that the impact of feedback on the shift is dependent on the applied shear strain. Most importantly, due to the shift in the transition, the system remains rigid (or stretching-dominated) when expected to be in a floppy state (or a bending-dominated state) in the tension-suppressed feedback limit. Understanding such a feedback mechanism can be instrumental in tuning the feedback for material designs that elicit a desired mechanical response.
\begin{figure}
    \centering
    \begin{minipage}{\columnwidth}
        \centering
      \includegraphics[height=0.95\columnwidth ,width=0.9\columnwidth]{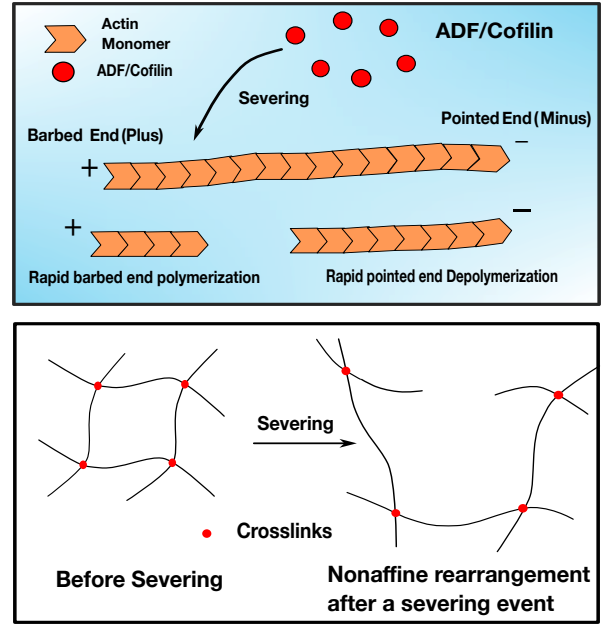}       %
      \caption{The first panel of the schematic demonstrates severing via cofilin in an actin filament. Severing occurs when cofilin binds to an actin filament, resulting in filaments with free barbed (plus) and pointed (minus) ends. These filaments can undergo rapid polymerization (at the barbed end) and depolymerization (at the pointed end) in the presence of ATP/ADP, depending on local actin monomer concentration. The second panel shows a severing event in the context of an actin network. Fragmentation results in dangling ends and nonaffine structural rearrangement due to relaxation.}
    \label{SeveringSchematic}
    \end{minipage}
\end{figure}
\section{MODEL}
We use 2D triangular lattice-based networks to investigate the effects of mechano-chemical feedback of stress on severing. These systems are commonly used to understand the rheological properties of biopolymer fiber networks \cite{arzash2019stress,arzash2020finite,broedersz2014modeling,shivers2019normal,shivers2019scaling,lee2022stiffening,sharma2016strain,arzash2022mechanics,gannavarapu2024effects}. 

We begin by generating networks on a two-dimensional periodic lattice with freely hinging crosslinks at the intersection points. The lattice spacing is 1 ($ l_{0} = 1$). A fully connected structure has a local connectivity ($z$) of 6, which exceeds the Maxwell isostatic threshold of 4 in 2D ($ z_{iso} = 2d$ for d dimensions). As noted above, in naturally occurring biopolymers like collagen and actin, the average local connectivity of these networks is below the isostatic threshold. To incorporate this, we lower the average connectivity via random dilution until it reaches a specified average $z<4$. We also cut one bond from each fiber randomly to eliminate the presence of any straight, system-spanning fibers that would lead to an unphysical mechanical stability in our finite simulations. 

In the cytoskeleton, actin networks are highly dynamic due to various reactions such as polymerization, depolymerization, severing, etc \cite{cooper1991role,pollard1986rate,pollard2003cellular}. Severing events lead to fragmentation of filaments, resulting in dangling ends in the network. A schematic of severing via cofilin binding for actin filaments and the consequential structural rearrangement in the network due to sudden loss of tension is depicted in \figurename{\ref{SeveringSchematic}}. Dangling ends have no impact on the mechanical behavior of the network due to quick relaxation and, therefore, are removed. To isolate the stress relaxation behavior via severing, we do not consider polymerization or depolymerization reactions at the dangling ends. 

Severing reactions occur throughout the network and are treated as independent events resulting in the release of tension with each occurrence. 
After each severing event, the network is relaxed to its minimum-energy state before the next event occurs. In this framework, there are three relevant timescales, the filament fragmentation timescale ($t_{\text{fragmentation}}$), the mechanical relaxation timescale of the network ($t_{\text{relaxation}}$) and the severing reaction timescale ($t_{\text{severing}}$) which is the time elapsed until the next event. We consider the limit in which fragmentation and network relaxation are both fast compared to the severing reaction timescale. Under this separation of timescales, the only timescale explicitly treated is $t_{\text{severing}}$, which is associated with the rate-dependent severing events. 
\begin{equation}
    t_{\text{fragmentation}} \ll t_{\text{relaxation}} \ll t_{\text{severing}}
\end{equation}

\begin{figure}
\centering
  \begin{subfigure}{0.475\textwidth}
  \includegraphics[height=0.6\textwidth]{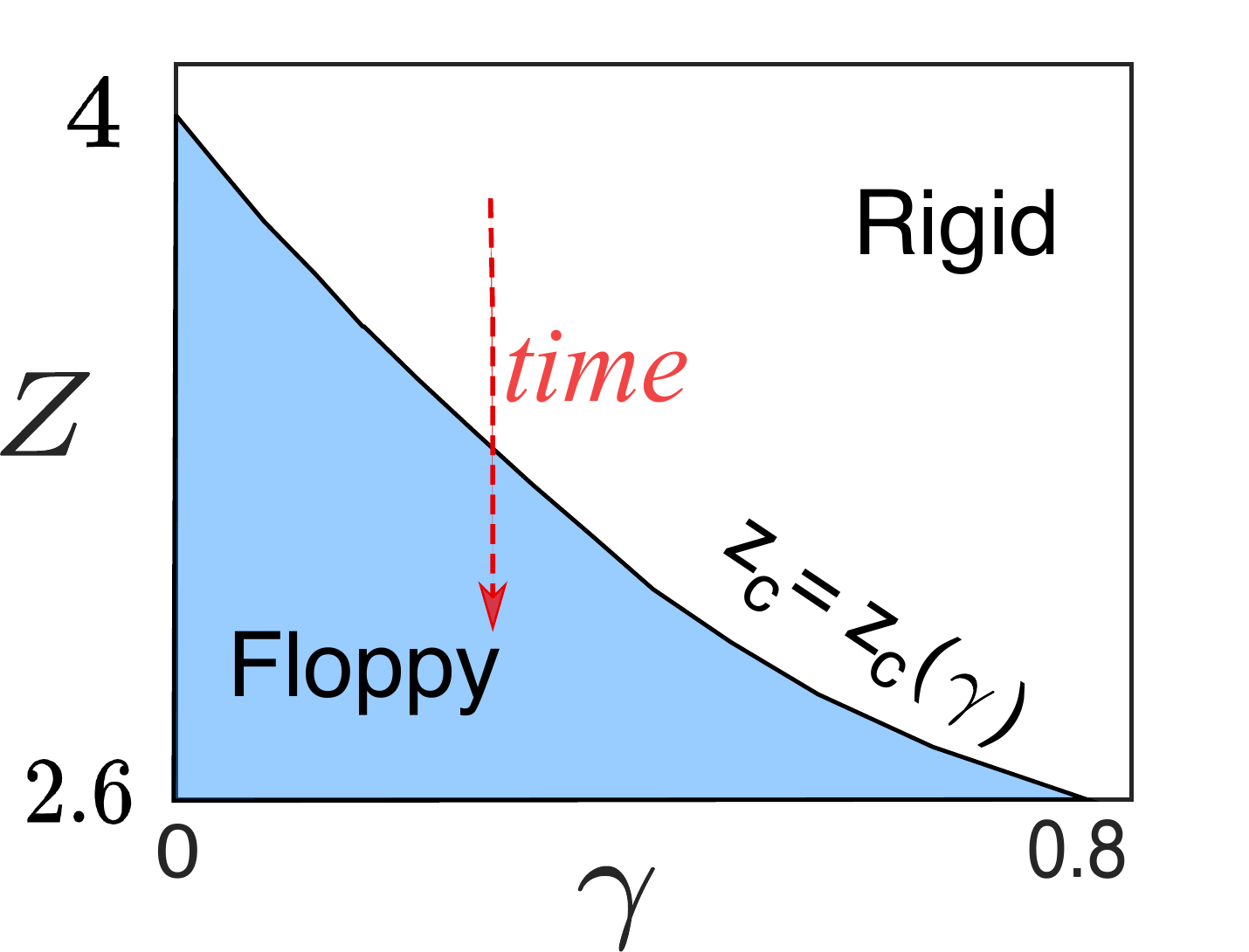}
  
  \caption{}
  \label{PhaseDiagSchematic_a}
  
\end{subfigure}\hfill 
\centering
\begin{subfigure}{0.475\textwidth}

\includegraphics[height=0.6\textwidth]{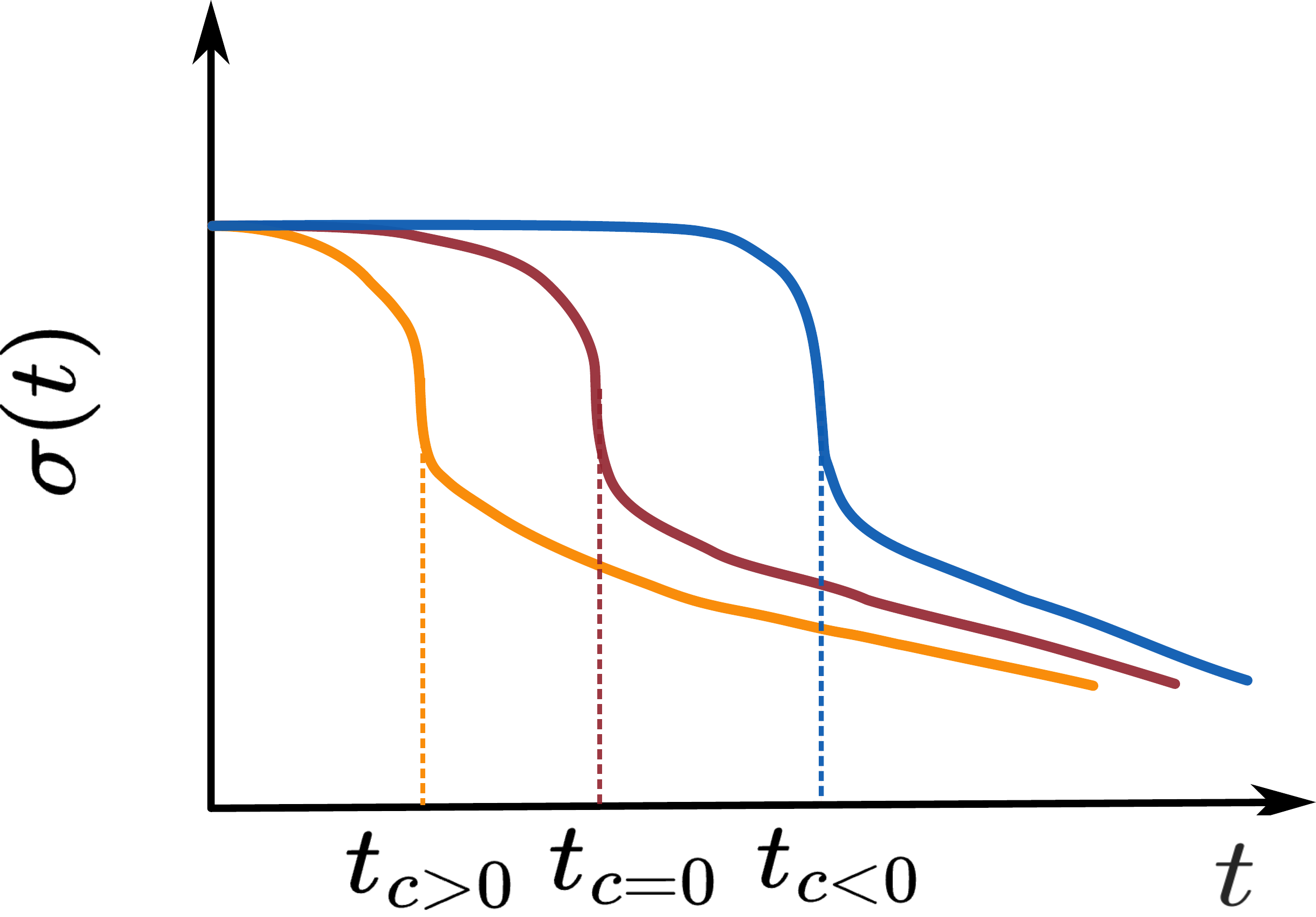}
  \caption{}
   \label{PhaseDiagSchematic_b}
\end{subfigure}

\medskip 
 \caption{(a) Diagram showing the mechanical phase transition in fiber networks. Networks with central force interactions undergo a mechanical phase transition controlled by the applied shear strain $ \gamma $ and average connectivity $ z $. In the absence of bending interactions, networks with a given connectivity $ z $ remain floppy when held under low applied shear strains and are rigid when held under high values of applied shear strains. With increasing applied strain, these networks exhibit a transition from a floppy to a rigid phase, and the transition, being critical in nature, can be depicted using the $ z_{c} = z_{c}(\gamma) $ curve in the figure. The mechanical transition is observed in the presence of bending interactions, but the floppy phase becomes bending-dominated instead. (b) A schematic plot of stress with time for depicting stress relaxation in three limits of feedback on severing: tension suppressed (blue line, $c<0$), tension-enhanced (orange line, $c>0$), and the no-feedback case (dark red line, $c=0$). The dotted lines indicate the corresponding times at which a dramatic stress drop occurs for each feedback limit.}
\end{figure}

After generating the network geometry, we treat network filaments between crosslinks as Hookean springs. In addition to the stretching energy of springs, we introduce additional bending interactions between nearest-neighbor filaments. The resulting Hamiltonian of the system includes stretching ($\mathcal{H_{\text{Stretch}}}$) and bending energies ($\mathcal{H_{\text{Bend}}}$) and can be written as

\begin{equation}
\mathcal{H_{\text{Total}}} = \mathcal{H_{\text{Stretch}}} + \mathcal{H_{\text{Bend}}}
\end{equation}
\begin{equation}
\mathcal{H_{\text{Total}}} = \frac{\mu}{2} \sum_{
\langle i,j \rangle} \frac{(l_{ij} - l_{ij,0})^2}{l_{ij,0}} + \frac{\kappa}{2} \sum_{ \langle i,j,k \rangle } \frac{(\theta_{ijk} - \theta_{ijk,0})^2}{\frac{1}{2}(l_{ij,0} + l_{jk,0})}
\end{equation}
where $\mu$ and $\kappa$ are stretching and bending stiffness, respectively. ${l_{ij}}$ and  $l_{ij,0}$ are current and initial/rest filament lengths between crosslinks $i$ and $j$, respectively, and $\theta_{ijk,0}$ and $\theta_{ijk}$ being the initial and current angles between neighboring filaments $ij$ and $jk$ respectively. We define dimensionless bending rigidity as $\tilde{\kappa} = \frac{\kappa}{\mu l_{0}^2}$ and set $\mu = 1$ as done in prior works \cite{arzash2020finite,broedersz2014modeling,shivers2019normal,shivers2019scaling,arzash2022mechanics,sharma2016strain,broedersz2011criticality}. 

We begin our simulations by applying a shear strain $\gamma$ and let severing events occur within the network. Between every two consecutive severing events, the network attains a minimum energy configuration. We use Lees-Edwards boundary conditions to apply shear strain as our system is periodic in both $x$ and $y$ directions. The network is held under the applied shear strain throughout the simulation, and we use the FIRE algorithm \cite{bitzek2006structural} to find the minimum energy configuration each time the system relaxes. We then calculate the stress tensor components as 

\begin{equation}
\sigma_{\alpha \beta} = \frac{1}{2A}\sum_{ij} f_{ij,\alpha} u_{ij,\beta}
\end{equation}
where $A$ is the simulation box area, $f_{ij,\alpha}$ is $\alpha$ component of the force exerted on crosslink $i$ from crosslink $j$, and $u_{ij,\beta}$ is the $\beta$ component of the displacement vector of the filament connecting the crosslinks $i$ and $j$. 

To model severing events in the network, we implement an event-driven simulation approach based on the Gillespie algorithm \cite{gillespie1977exact}. Severing events are treated as discrete, randomly occurring reactions for every filament with a propensity that represents the reaction rate $\alpha$. We determine the reaction rate ($\alpha_{ij}$) for every filament based on its current state of tension and calculate the total reaction propensity given by $\alpha_{\textrm{\textbf{total}}} = \sum_{ij} \alpha_{ij}$. Two random numbers $r_{1}$ and $r_{2}$ are generated to determine the time interval and the specific filament to be severed. The time interval for the next event is given by $\Delta t = -\frac{\ln(r_{1}) }{\alpha_{\textrm{\textbf{total}}}}$. To identify the location of the next severing event, we normalize the propensities and construct a cumulative distribution of probabilities ($p_{ij} = \frac{\alpha_{ij}}{\alpha_{\textrm{\textbf{total}}}}$). The next severing event location is identified by comparing the cumulative distribution of the propensities with $r_{2}$. Based on where $r_{2}$ falls in the cumulative distribution, the corresponding specific filament is chosen to be severed in the next event. The network state is then updated to reflect the severing event, and the simulation time is incremented by $\Delta t$.  Such mechano-chemical feedback can be introduced by implementing a modified severing rate given as follows
\begin{equation}
\mathlarger{\alpha = }
\begin{cases}
\mathlarger{\alpha_{0}} \ \mathlarger{{\rm e}}^{\mathlarger{ c  \tau}} & \mbox{when under tension } (\tau  \geq 0) \\
\mathlarger{\alpha_{0}}  & \mbox{when  under compression }     
\end{cases}
\end{equation}%
$\tau$ is the tension in the filament between crosslinks, and $c$ is the severing feedback parameter. %
  We analyze two feedback limits, tension-suppressed severing (negative $c$) and tension-enhanced severing (positive $c$), along with the no-feedback case. In the  ``tension-suppressed limit of severing", the rate decreases when under extension and remains unchanged during compression. In the other limit of the ``tension-enhanced limit of severing", the rate increases with tension and remains the same when under compression. A simplified schematic of stress relaxation in these three limits is shown in \figurename{\ref{PhaseDiagSchematic_b}}.

It has been well-established and extensively studied that sub-isostatic fiber networks ($z< z_{iso}$) undergo a floppy to rigid transition at a finite critical strain for a given connectivity. When held under low shear, the network remains floppy (bending-dominated in the presence of bending interactions). The stretching modes dominate at sufficiently large shear strains depending on the connectivity and the network exhibits rigid behavior. This strain-controlled transition is not only critical in nature but also is a function of network connectivity and geometry \cite{licup2015stress,sharma2016strain}. 
The phase boundary for this strain-controlled critical transition is shown in \figurename{\ref{PhaseDiagSchematic_a}}. The critical strain and connectivity can be expressed as $\gamma_{c} = \gamma_{c}(z)$ for a specified average network connectivity $z$ or $z_{c} = z_{c}(\gamma)$ for an applied shear strain $\gamma$ using the phase boundary. In our simulations, applied shear strain remains constant, and network connectivity changes as severing occurs. We investigate the change in the mechanical behavior of networks with connectivity for varying values of applied shear strain $\gamma$, dimensionless bending rigidity $\tilde{\kappa}$, and the tension feedback parameter $c$. %
\begin{figure}
\centering
  \begin{subfigure}{0.475\textwidth}
  \centering \includegraphics[height=0.8\textwidth,width=0.914\columnwidth]{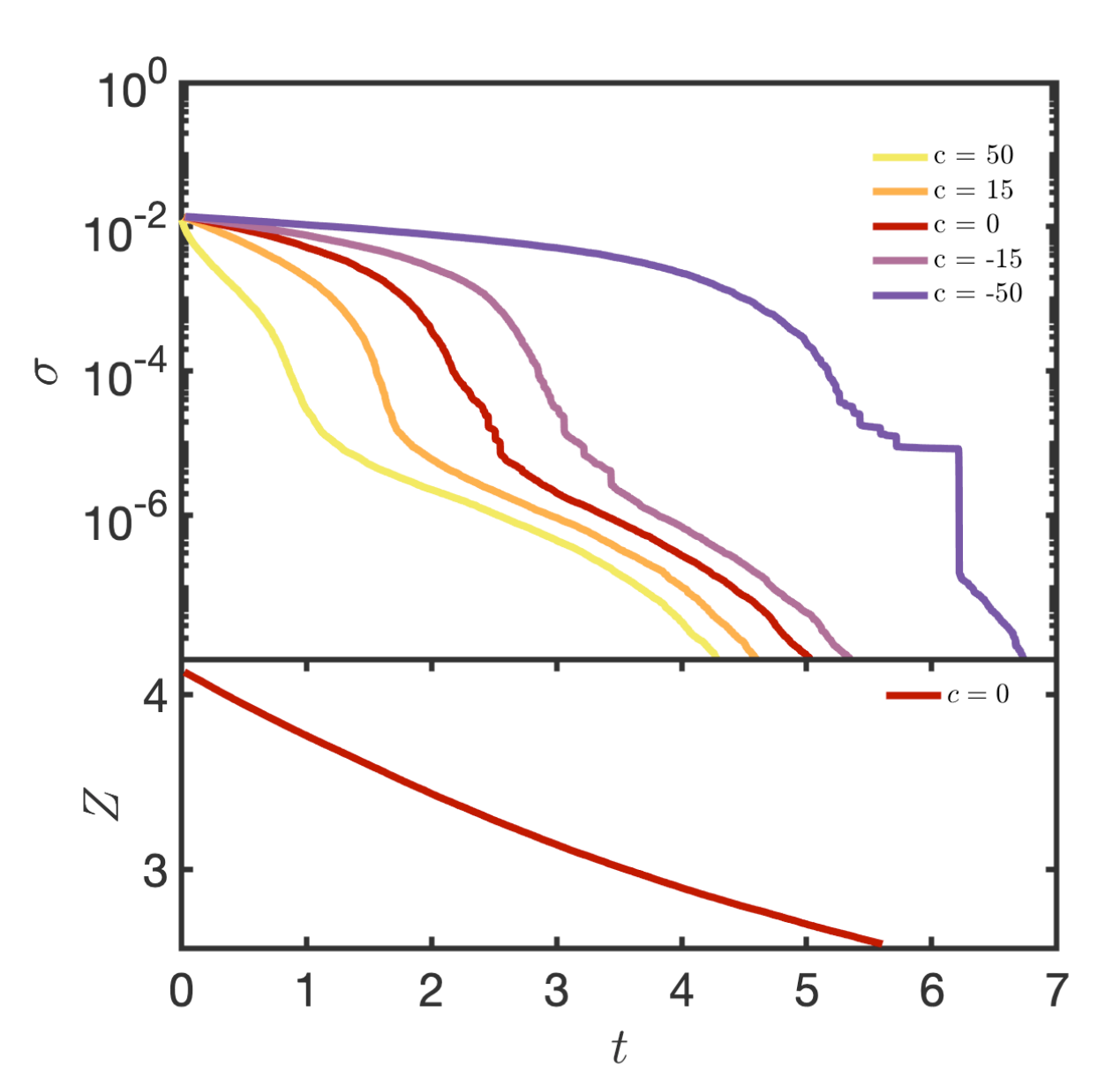} 
  \caption{}     \label{StressZVsTime_constStrain_3a}
\end{subfigure}\hfill 
\centering
\begin{subfigure}{0.475\textwidth}
\centering
  \includegraphics[height=0.8\textwidth,width=0.914\columnwidth]{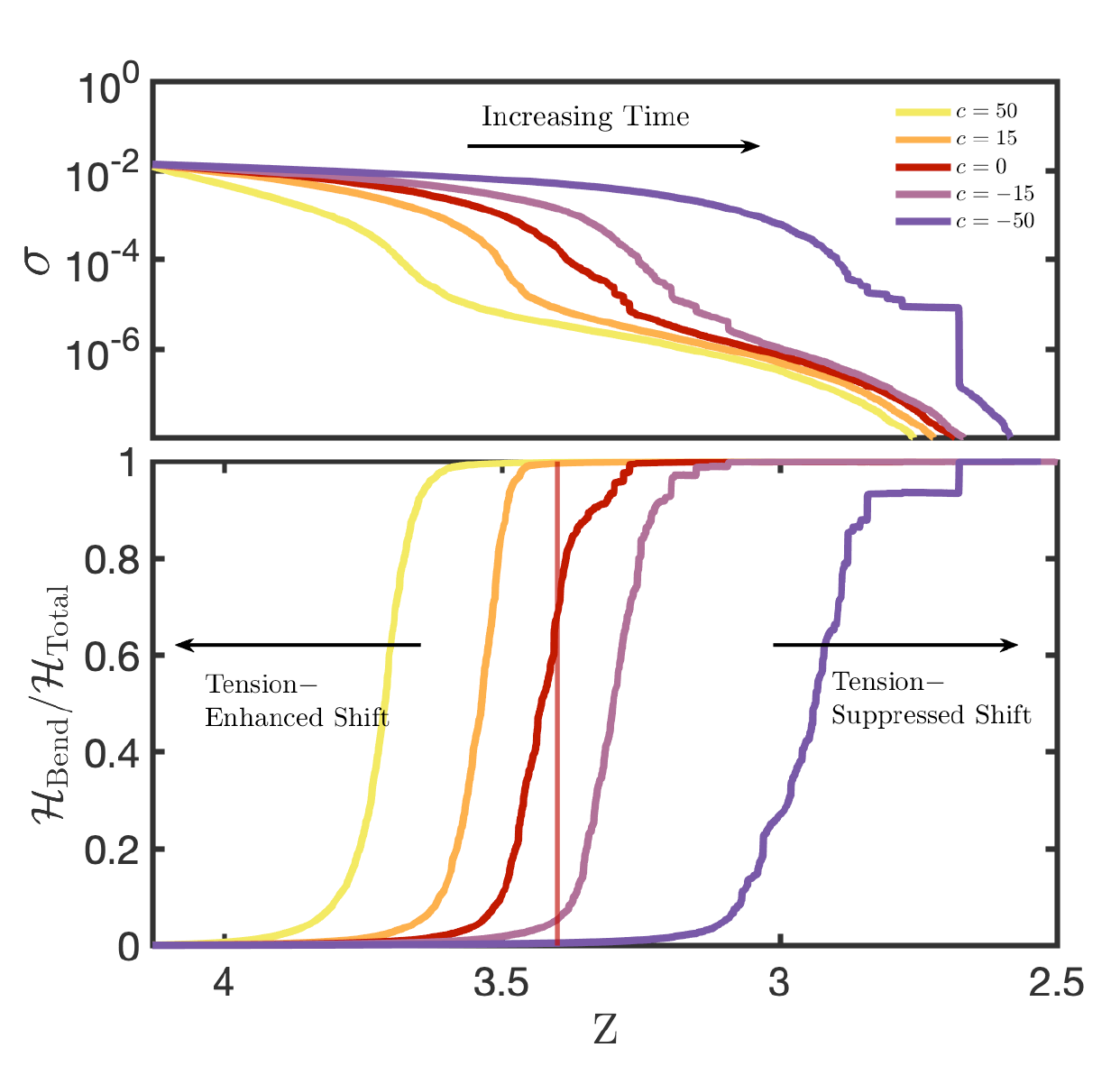}
  \caption{}
   \label{StressEnergyVsZ_constStrain_3b}
\end{subfigure}
\medskip %
\caption{(a) Stress relaxation $\sigma $ and Average connectivity $z$ as a function of time $t$ for varying values of the tension feedback parameter (positive, negative, and zero). (b) Variation of Stress $\sigma$ and ratio of Bending energy $\mathcal{H_{\text{Bend}}}$ to Total energy $\mathcal{H_{\text{Total}}}$ with Connectivity $z$ for the networks. The networks have a dimensionless bending rigidity of $\tilde{\kappa} = 10^{-5}$ and are held under an applied shear strain $ \gamma = 0.2.$ From Fig. \ref{StressZVsTime_constStrain_3a}, it can be observed that connectivity $z$ decreases with time due to the occurrence of severing events. For better physical understanding, stress $\sigma$ can be analyzed with connectivity $z$, as shown in Fig. \ref{StressEnergyVsZ_constStrain_3b}, for the same networks. Connectivity $z$ is plotted on the horizontal axis decreasing from left to right. The arrow indicates the direction of increasing time in accordance with the decrease in connectivity with time, as shown in Fig \ref{StressZVsTime_constStrain_3a}. The bold vertical line denotes the expected connectivity at transition for the applied shear strain $\gamma = 0.2$, and the simulation results also exhibit phase transition at that connectivity in the absence of feedback.
}
\label{StressEnergyZTimeconstStrain}
\end{figure}
\section{RESULTS}
The mechanical response of fiber networks has been explored before for various network geometries using the standard method of applying incremental deformations \cite{arzash2020finite,shivers2019scaling,arzash2022mechanics,chen2023effective,lee2022stiffening}. To confirm the behavior, we first discuss the variation of relevant system parameters to 2D triangular networks with connectivity $z$ at a constant applied shear strain $\gamma$. Before introducing finite mechano-chemical feedback, we consider the case where feedback is absent, i.e., $c  = 0$. When feedback parameter $c = 0$, severing rates corresponding to each filament are constant, i.e., $\alpha_{ij} = \alpha_{0}$ regardless of filament state of extension or compression. With all propensities being the same, this case of severing corresponds to random dilution. Starting with a specified connectivity under an applied shear strain, we expect the system to behave as a standard 2D triangular lattice-based network of springs. \figurename{\ref{StressEnergyZTimeconstStrain}} shows the stress relaxation $\sigma(t)$ and variation of average network connectivity $z$ with time $t$ for different feedback parameters for dimensionless bending rigidity $\tilde{\kappa} = 10^{-5}$. Severing events result in tension loss and filament deletion. Therefore, the average connectivity $z$ of the network decreases as severing occurs with time. We find analyzing the mechanical properties with respect to the average connectivity more insightful than that with time, as the transition is connectivity and strain-controlled. 

To understand if relaxation occurs through stretching or bending modes, we plot the ratio of fiber-bending energy to the total energy $(\mathcal{H_{\text{Bend}}}/\mathcal{H_{\text{Total}}})$ for the same networks as shown in the figure. The arrow serves to indicate the direction of increasing time when displaying the variation of relevant quantities with connectivity on the X-axis, illustrating the decrease in connectivity over time. Starting with super-isostatic networks, we expect the stretching modes to dominate. Severing events occur at random locations in the network, resulting in stress relaxation. Using the phase boundary, we calculate the expected critical connectivity $z_{c}$ for the applied shear strain $\gamma$ using $z_{c}= z_{c}(\gamma)$ from the phase boundary. The network remains in the stretching-dominated regime for $z>z_{c}$ until it transitions to a bending-dominated regime as the connectivity drops to $z_{c}$. The data shown in Fig. \ref{StressEnergyZTimeconstStrain} are for a dimensionless bending rigidity $\tilde{\kappa} = 10^{-5}$. In the absence of feedback, the transition occurs at the expected critical connectivity, denoted by the bold vertical line for the applied shear strain. In presence of feedback, we observe a shift towards lower or higher connectivities depending on the feedback. The shift is dependent on applied strain as well as the feedback parameter.

 As the connectivity drops below the strain-dependent critical value, a sharp drop is observed in stress levels, particularly evident at low dimensionless bending rigidities, as depicted in Fig.\ref{StressEnergyVsZ_constStrain_3b}. Each severing event results in a sudden loss of tension due to filament fragmentation. Triangular networks form branch-like force chains near the transition, crucial for rigidifying the otherwise floppy network. 
 The dramatic stress drop at the critical point occurs as load-bearing filaments start breaking. As the network becomes sparser, displacement fluctuations increase to accommodate the applied shear strain.
 \begin{figure}
\begin{subfigure}{0.475\textwidth}
\includegraphics[height=0.8\textwidth,width=0.914\columnwidth]{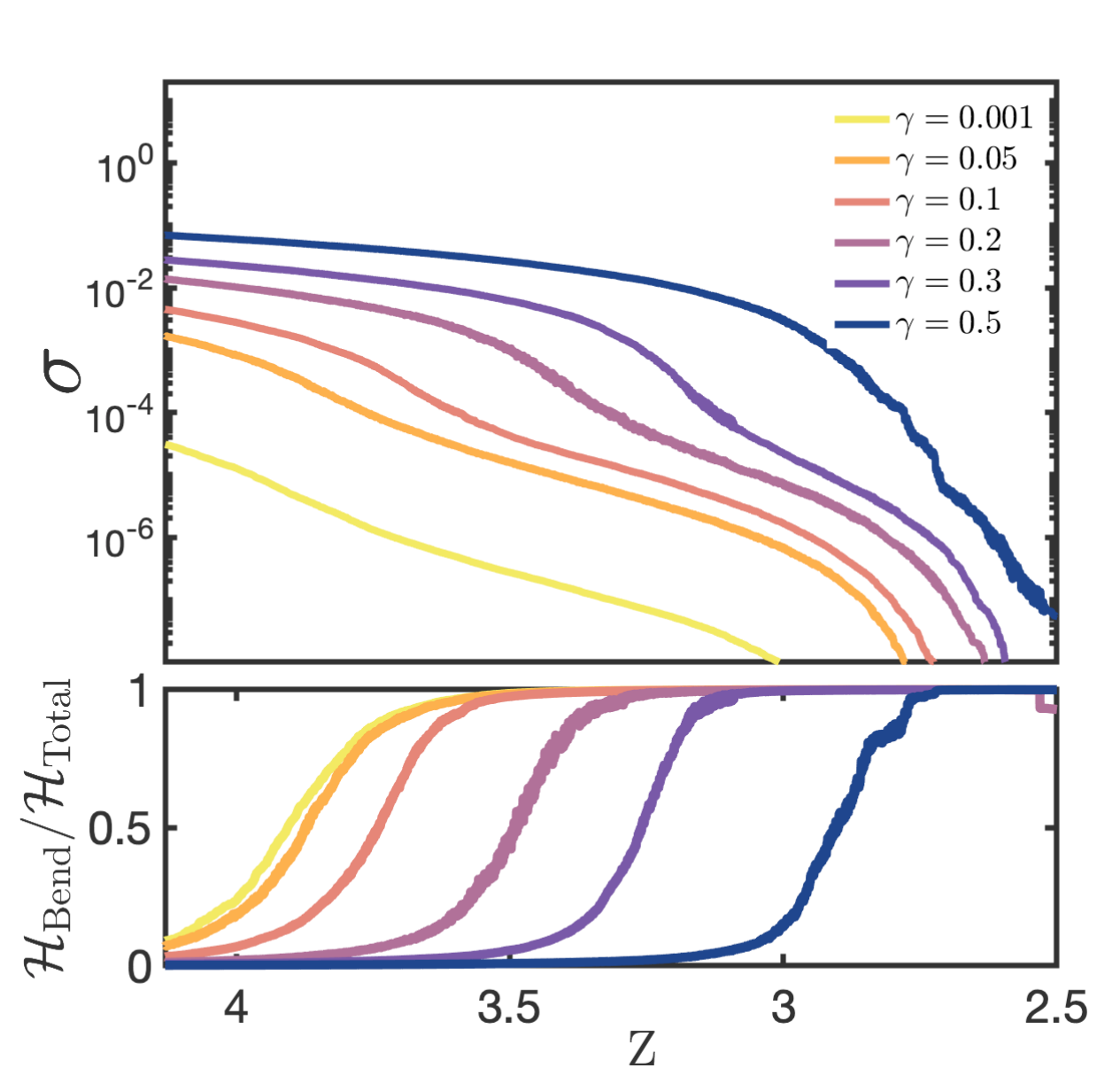}

  \caption{}
       \label{StressEnergyVsZ_Kappa1e-4}
\end{subfigure}\hfill 
\begin{subfigure}{0.475\textwidth}
\includegraphics[height=0.8\textwidth,width=0.914\columnwidth]{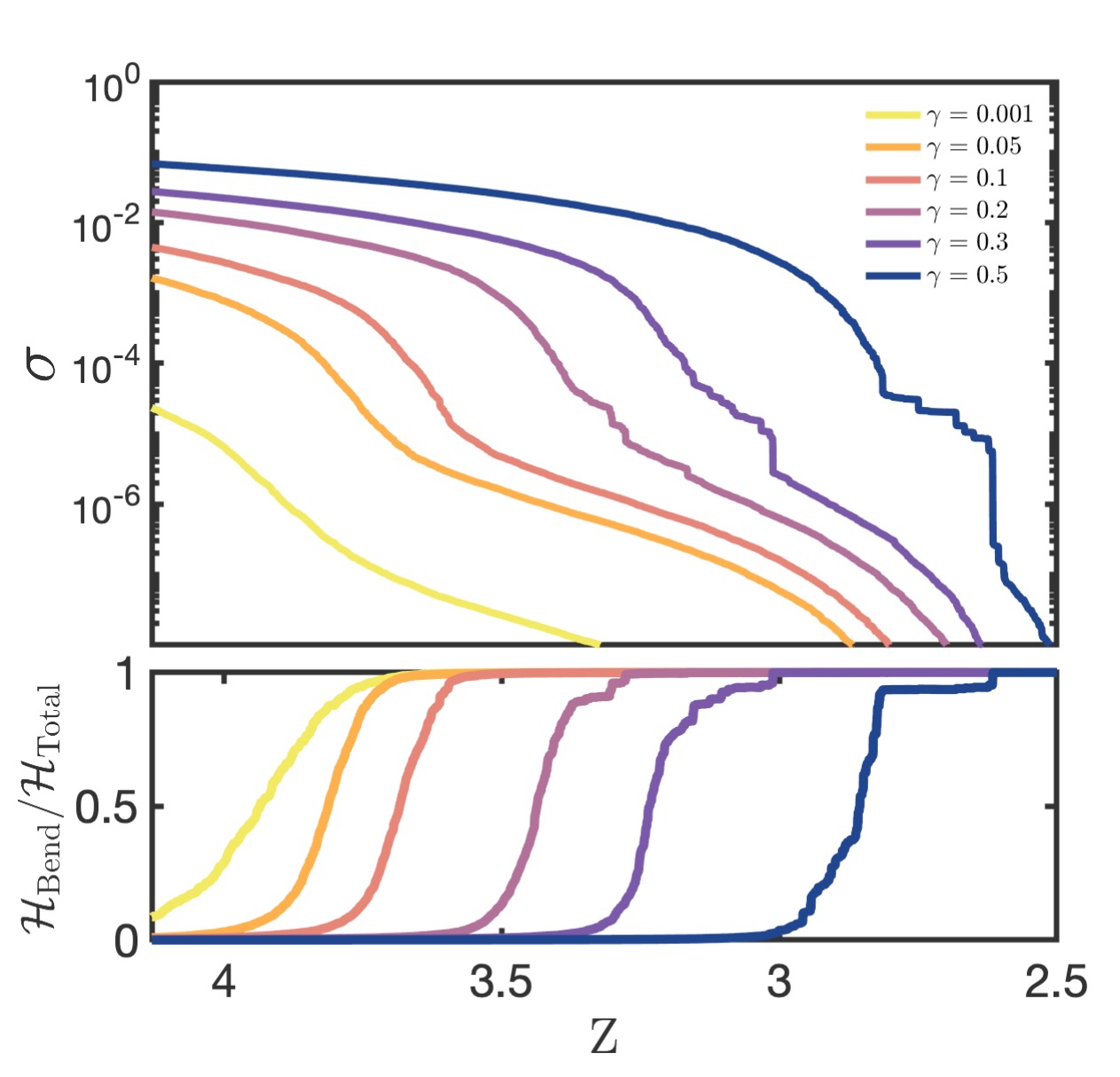}

  \caption{}
     \label{StressEnergyVsZ_Kappa1e-5}
\end{subfigure}
\caption{Absence of feedback ($c=0$): Stress $\sigma$ and ratio of Bending energy $\mathcal{H_{\text{Bend}}}$ to Total energy $\mathcal{H_{\text{Total}}}$ with Connectivity $z$ for the networks as a function of connectivity $z$ for varying applied strain $\gamma$. Fig \ref{StressEnergyVsZ_Kappa1e-4} and \ref{StressEnergyVsZ_Kappa1e-5} correspond to behavior of networks with dimensionless bending rigidity of $\tilde{\kappa} = 10^{-4}$ and $10^{-5}$ respectively with no feedback present. Bending rigidity clearly affects the behavior near the phase transition, with higher bending rigidity making it more gradual. This is in agreement with fiber network behavior.}
     \label{StressEnergyVsZ_Kappa}
\end{figure}%

 Now, we introduce finite mechanochemical feedback of stress on severing. We replicate the analysis conducted in the absence of feedback to show the impact of feedback on the phase transition. In the presence of feedback, filament severing rates are subjected to change with tension. Specifically, in the tension-suppressed severing scenario ($c<0$), filaments experiencing higher tension exhibit a diminished propensity for severing compared to those under compression or lower tension. Conversely, filaments subjected to higher tension are more prone to severing in the tension-enhanced severing scenario ($c>0$). Consequently, the severing rate distribution across filaments becomes asymmetric: while compressed filaments remain unaffected by the feedback, severing rates either increase or decrease for filaments experiencing tension, based on the feedback limit. 
 
 To assess the influence of feedback on the transition, we analyze stress relaxation $\sigma(t)$ and the variation of average connectivity $Z$ with time for different feedback values under a constant applied shear strain $\gamma$.
 As depicted in Fig. \ref{StressZVsTime_constStrain_3a}, the dramatic stress drop expected at the onset of rigidity occurs at further times for tension-suppressed severing and earlier for tension-enhanced severing limits. This happens due to lower severing rates observed for filaments experiencing high tension, resulting in a preference for severing filaments under lower tension in the tension-suppressed limit. In the tension-enhanced limit, it occurs earlier owing to higher severing rates for high-tension filaments.
 
 Additionally, we plot the ratio of fiber-bending energy to the total energy $(\mathcal{H_{\text{Bend}}}/\mathcal{H_{\text{Total}}})$ for different $ c $ values at a constant $\tilde{\kappa}$ as shown in Fig.\ref{StressEnergyVsZ_constStrain_3b}. 
 The bold vertical line signifies the anticipated connectivity at the transition. The $c=0$ behavior aptly aligns in accordance with the expected $z_{c}$. The energy ratio curves clearly show the shift in the transition due to the feedback. The transition from stretching-dominated to bending-dominated regime occurs at connectivities lower than the critical connectivity for applied strain in the tension-suppressed severing limit. Moreover, the shift towards lower connectivities depends on the feedback parameter's magnitude. The higher the magnitude of feedback in tension-suppressed limit, the more the shift towards lower connectivities.

 To gain better clarity,  we analyze the stress relaxation with connectivity in Fig. \ref{StressEnergyVsZ_constStrain_3b}, which depicts the dramatic drop in stress. In the absence of feedback, it occurs at the anticipated connectivity consistent with the transition. However, a shift is observed in the presence of feedback. This shift can be attributed to the severing of load-bearing network filaments. When feedback is absent, the force chains are equally likely to be severed as any other filament. However, tension-suppressed feedback biases against severing these force chains, delaying the stress drop until severing occurs at load-bearing filament locations. Conversely, tension-enhanced severing leads to stress drop at relatively higher connectivities than expected.

Networks (without bending interactions) below the critical point are floppy and can be stabilized by adding weak bending interactions or prestress. Additional bending modes introduce a finite stress (and hence a modulus) in the linear regime, stabilizing an otherwise floppy system. When held under shear strain, the network relaxes to a nonaffine minimum-energy configuration to avoid the energetically costly configuration associated with affine deformation.
After each severing event, the network undergoes local structural rearrangements to reach the energetically favorable configuration. 
To quantify these rearrangements, we calculate the nonaffinity of the network given by
\begin{equation}
    \mathlarger{\Gamma = \frac{\langle || \delta u^{\textbf{NA}} ||^2 \rangle }{l_{0}^2 \gamma^2}}
\end{equation}
where $l_{0}$ is the initial filament length and $\delta u^{\textbf{NA}} = \textbf{u} - \textbf{u}^{\textbf{affine}}$. Here, $\textbf{u}$ represents the current position of the node, and $\textbf{u}^{\textbf{affine}}$ is the position of the node corresponding to an affine deformation for strain $\gamma$. This quantity is averaged for all the network nodes at each simulation step. As established in prior works, these $\tilde{\kappa}$ dependent nonaffine fluctuations exhibit a peak at the critical strain when the system is subjected to incremental shear deformation. We also investigate how the nonaffine fluctuations change with the network connectivity maintained at a constant shear strain.

Fig.\ref{StressEnergyVsZ_Kappa} illustrates the network behavior at two different values of bending rigidities, with all other parameters held constant. We analyze the behavior by varying shear strains for dimensionless bending rigidities of $\tilde{\kappa} = 10^{-4}$ and $10^{-5}$ in the absence of feedback, i.e., $c=0$. 
Bending rigidity does not affect the onset of transition but at higher bending rigidity, the transition is more gradual.
This effect is further observed in stress relaxation, where a dramatic stress drop occurs at the anticipated connectivity at transition. The findings presented confirm that the mechanical behavior in the absence of feedback aligns with prior observations of fiber network behavior under shear.
 \begin{figure}
\begin{subfigure}{0.475\textwidth}
  \includegraphics[height=0.8\textwidth,width=\columnwidth]{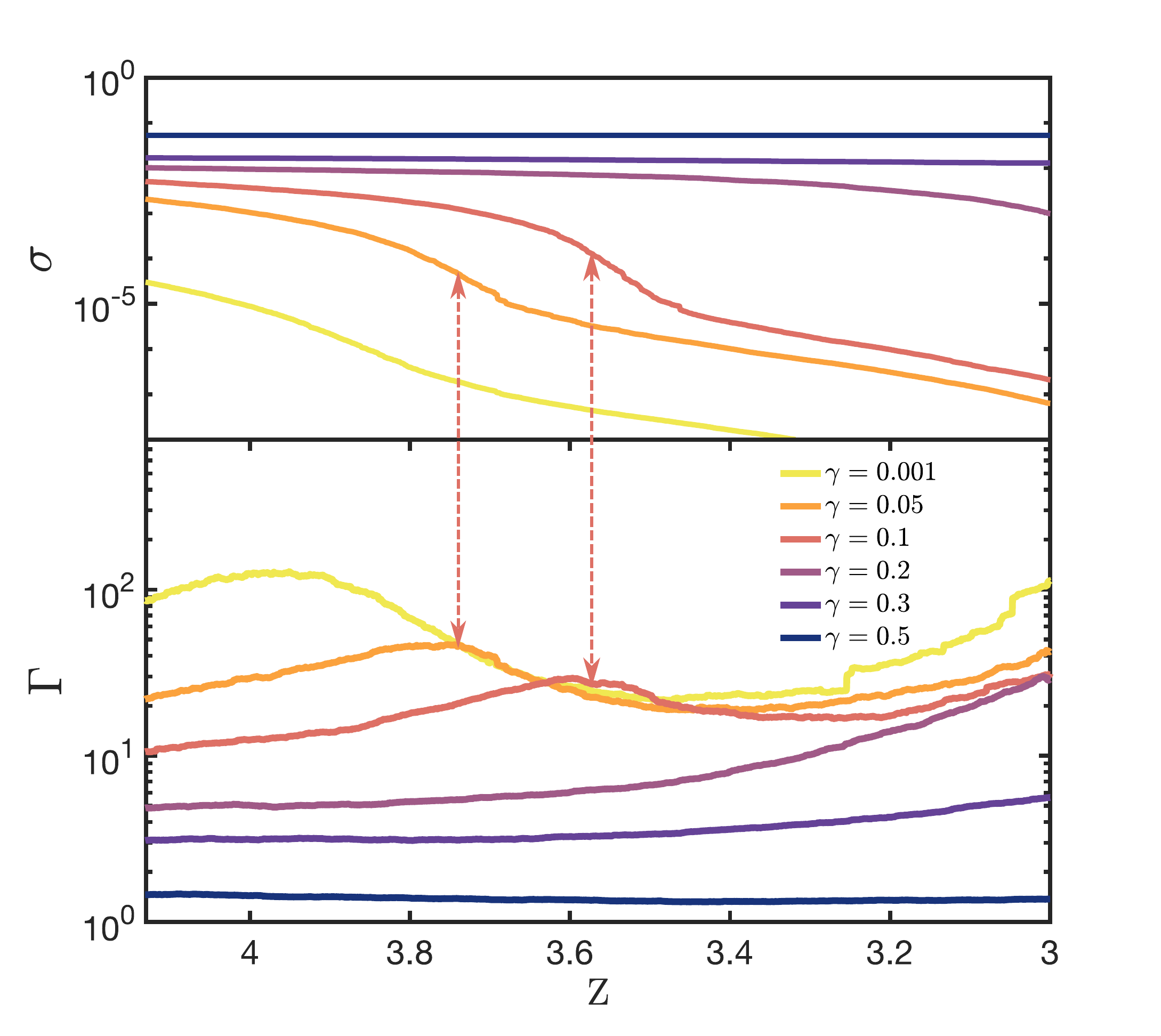}
  
  \caption{}
  
       \label{StressEnergyVsZ_Param-50}
\end{subfigure}\hfill 
\begin{subfigure}{0.475\textwidth}
  \includegraphics[height=0.8\textwidth,width=\columnwidth]{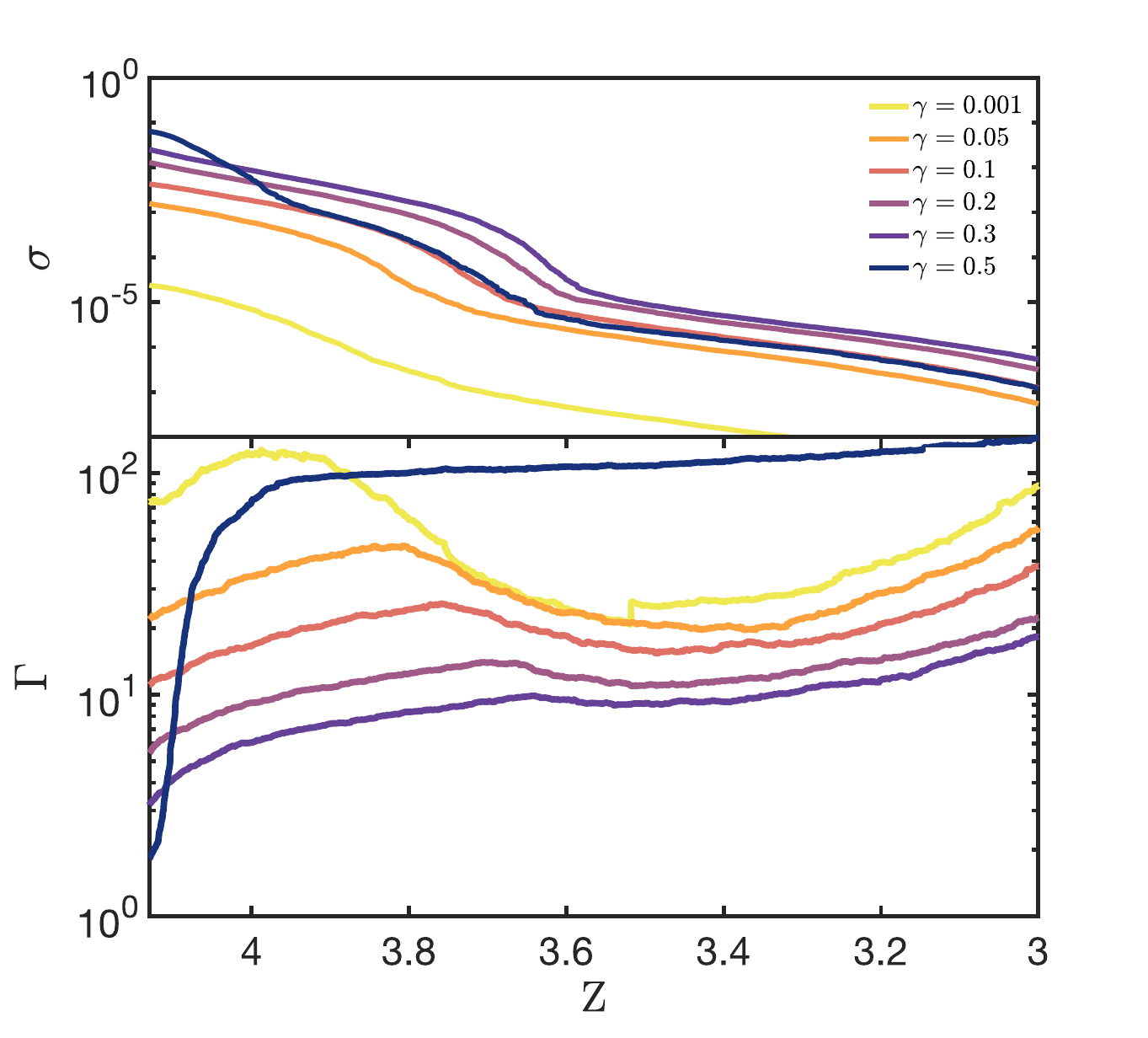}
  \caption{}
         \label{StressEnergyVsZ_Param50}
\end{subfigure}
\caption{Presence of Feedback: Stress $\sigma$ and Nonaffinity $\Gamma$ with connectivity $z$ for the networks for varying applied strain $\gamma$. The top and bottom panels correspond to the behavior of networks with tension feedback parameter values of $c = -50$  and $c = 50$, respectively. For smaller strains, the behavior is similar to when the feedback is absent. In the presence of moderately high strains, a clear shift in the transition can be observed in both tension-suppressed ($c = -50$) and tension-enhanced ($c = 50$) severing cases. The ``shift" in the transition is dependent on the applied strain. The dramatic stress-drop aligns with a peak in Nonaffinity indicating the transition into bending-dominated regime. } 

       \label{StressEnergyVsZ_constParam}
\end{figure}

We now study the effect of the strength of tension feedback by examining stress relaxation behavior for different values of applied shear strain $\gamma$ for a given bending rigidity. In Fig. \ref{StressEnergyVsZ_Param-50} and \ref{StressEnergyVsZ_Param50}, we plot stress ($\sigma$) and Nonaffinity ($\Gamma$) vs connectivity ($z$) for several values of shear strain ($\gamma$). The loss of rigidity can be identified by a sharp drop in stress ($\sigma$) and a peak in nonaffinity is observed that aligns with the drop as shown in Fig. \ref{StressEnergyVsZ_Param-50}. 

The effect of the same feedback parameter results in a greater shift for higher strains than for lower strains. This behavior arises because the strength of feedback is not only dependent on the magnitude of the feedback parameter but also on the filament tensions generated by the applied deformation. At a fixed value of the feedback parameter, filaments experience larger tensions at larger values of shear strain ($\gamma$). In the tension-suppressed limit, filaments with high tension are more likely to remain intact whereas in the tension-enhanced limit, they are preferentially severed. This indicates that the breakdown of load-bearing chains is influenced by both applied shear strain and the feedback parameter. 

In tension-suppressed severing, the bias against filaments under high tension amplifies with increasing applied shear strain. This allows load-bearing chains to persist to lower connectivities, shifting the transition toward smaller $z$. This trend is evident in Fig. \ref{StressEnergyVsZ_Param-50}. The behavior at low strains resembles the no-feedback case, while at high strains, the shift is high enough that the transition is not observed within the considered range of connectivities. 

 In Fig.\ref{StressEnergyVsZ_Param50}, the tension-enhanced case shows similar trends in the opposite direction, with the transition shifting towards higher connectivity. For lower strains, there is no significant deviation from the no-feedback case. However, at larger applied strains, filaments under high tension are preferentially severed. This results in load-bearing filaments breaking down and the transition into the bending-dominated regime occurring at higher connectivities. 
 
These findings suggest that the influence of tension-dependent severing is controlled by both the feedback parameter and the applied shear strain. For moderate to high applied shear strains, higher filament tensions result in a larger shift in the transition depending on the feedback limit. At low shear strains, feedback has weaker effects unless the feedback parameter is sufficiently large.

\section{SUMMARY AND DISCUSSION}
In our work, we have demonstrated stress relaxation in spring networks via severing in absence and presence of a finite mechano-chemical feedback of stress on severing. Our model explores two limits of the feedback, tension-suppressed and tension-enhanced severing. Under separation of timescales, network relaxation and filament fragmentation are fast compared to severing. In the absence of feedback, stress relaxation behavior via severing is consistent with prior observations of the mechanical phase transition of fiber networks under shear. For a fixed applied strain, severing progressively reduces the network connectivity. The system remains rigid (stretching-dominated regime) until the connectivity drops below the expected critical value of connectivity given by the phase diagram. Beyond this, the network crosses over into the floppy (bending-dominated) regime. Hence, we observe the dominance of stretching and bending modes at higher and lower connectivities, respectively, and the dramatic stress drop marks the phase transition.

In contrast, the addition of the mechanochemical feedback clearly modifies the expected behavior depending on the feedback limit. This feedback introduces a bias in filament selection for fragmentation according to the tension, and results in a shift in the transition. This shift is observed in both tension-suppressed and tension-enhanced severing limits, but in opposite directions. In the tension-suppressed limit, filaments with high tension are less likely to sever, which delays the rigidity loss and the transition occurs at connectivities lower than those predicted from the phase diagram. As a result, the networks maintain rigidity in a regime where they would otherwise be floppy until the connectivity decreases further with severing. In the tension-enhanced limit, however, the filaments with high tension are more likely to sever, which accelerates the rigidity loss, and the transition into bending dominated regime occurs earlier at higher connectivities than expected.  While emphasizing the evident transition shift, we elucidate how the applied shear strain and the magnitude of the feedback parameter influence this feedback's impact. 
The feedback is not significant for low-shear strains unless the feedback parameter is very high. However, for moderate to high strains, it depends on the magnitude of the feedback parameter.

Our work sheds light on how tuning feedback mechanisms can be extremely valuable in designing networks that behave according to requirements. Understanding such mechanochemical feedback can be utilized in designing networks that maintain rigidity in conditions when expected to be floppy \cite{wang2025mechanosensitive}. While our simulations primarily focus on shear strains, exploring bulk strains and their interaction with feedback on filament deletion could provide further insights.  Overall, our work contributes to understanding the intricate behavior of biological networks and may inspire experimental investigations in this domain.

\begin{acknowledgments}
The work was supported in part by the Center for Theoretical Biological Physics sponsored by the NSF (PHY-2019745). PK and FCM were also supported in part by the National Science Foundation Division of Materials Research (Grant No. DMR-2224030).
ABK also acknowledges the support from the Welch Foundation (C-1559).
\end{acknowledgments}

\bibliography{Version_1.0}

\appendix

\end{document}